\documentclass{sig-alternate-05-2015}

\usepackage{array}
\usepackage{multirow}
\newcolumntype{C}[1]{>{\centering\arraybackslash}p{#1}}

\def\sharedaffiliation{%
\end{tabular}
\begin{tabular}{c}}

\begin{document}

\setcopyright{rightsretained}





%

\title{Configurable memory systems for embedded many-core processors}
%
%
%
%
%

\numberofauthors{3} 
%
%
    \author{
      \alignauthor Daniel Bates\\      
      \alignauthor Alex Chadwick\\     
      \alignauthor Robert Mullins\\    
      \sharedaffiliation
       \affaddr{Computer Laboratory}\\
       \affaddr{University of Cambridge}\\
       \affaddr{UK}\\
       \email{\large\sffamily FirstName.LastName@cl.cam.ac.uk}\\
    }

\date{11 December 2015}

\maketitle
\begin{abstract}
The memory system of a modern embedded processor consumes a large fraction of total system energy. We explore a range of different configuration options and show that a reconfigurable design can make better use of the resources available to it than any fixed implementation, and provide large improvements in both performance and energy consumption. Reconfigurability becomes increasingly useful as resources become more constrained, so is particularly relevant in the embedded space.

For an optimised architectural configuration, we show that a configurable cache system performs an average of 20\% (maximum 70\%) better than the best fixed implementation when two programs are competing for the same resources, and reduces cache miss rate by an average of 70\% (maximum 90\%). We then present a case study of AES encryption and decryption, and find that a custom memory configuration can almost double performance, with further benefits being achieved by specialising the task of each core when parallelising the program.
\end{abstract}

%
%


%
%

%
%



\section{Introduction}
Consumers expect a relentless increase in performance of their mobile devices while also demanding longer battery lives. Since battery capacities are increasing only slowly, these requirements combine such that a huge improvement in energy efficiency is necessary. There is a precedent for this kind of progress, but many factors now conspire to slow this development to a halt. Virtually all modern consumer devices are thermally limited, so cannot run at a higher power to compute more quickly. Dennard scaling has slowed significantly, along with the energy efficiency improvements it brings. While improvements such as larger caches and wider-issue pipelines can improve performance, that improvement diminishes as more resources are spent, and power consumption tends to increase. Transistor scaling as dictated by Moore's Law is also showing signs of slowing, meaning we will need to use the same transistors for more design iterations, and therefore use them more intelligently.

All of these factors point towards a need for specialisation, and indeed, this is a direction that processor architects have been exploring lately. Specialisation allows unnecessary resources to be pruned away and specific optimisations to be applied without hurting the common case, leaving less area and energy overhead and improved performance. This approach of application-specific design has its own pitfalls, however, so we believe that the current trend is only a temporary deviation from the long-term pattern. Hardware specialisation adds to the already ballooning NRE costs involved in chip manufacture, since those costs now apply to a more restricted design, and more designs are needed to cover a general-purpose set of applications. There have been no recent breakthroughs in the productivity of microarchitects, so a more-complex design takes longer to build, and risks being obsolete by the time it is complete. We are entering an era of unreliable transistors, and having a large number of unique designs means that each one must be addressed individually to protect it from faults, increasing complexity further. The costs of testing and verification are also increasing super-linearly, and will only continue to do so if we continue along this path.

For the reasons outlined above, we advocate the use of software specialisation, where a simple, homogeneous fabric is used as a substrate for a specialised software overlay which can be reconfigured rapidly at runtime. By creating this substrate from a small selection of simple building blocks, we sidestep the problems of increased complexity, while also improving energy efficiency by eliminating hardware overheads. Depending on the type of computation required, we expect the design of this substrate to lie somewhere between an FPGA and a multi/many core processor. Such an architecture will consist of a large number of identical elements capable of computation, and interconnection to allow them to communicate efficiently.

In this paper, we focus on the memory system and explore ways in which a homogeneous architecture can be effectively reconfigured to improve efficiency.

\section{Loki}
\begin{figure}
  \centering
  \includegraphics[width=\columnwidth]{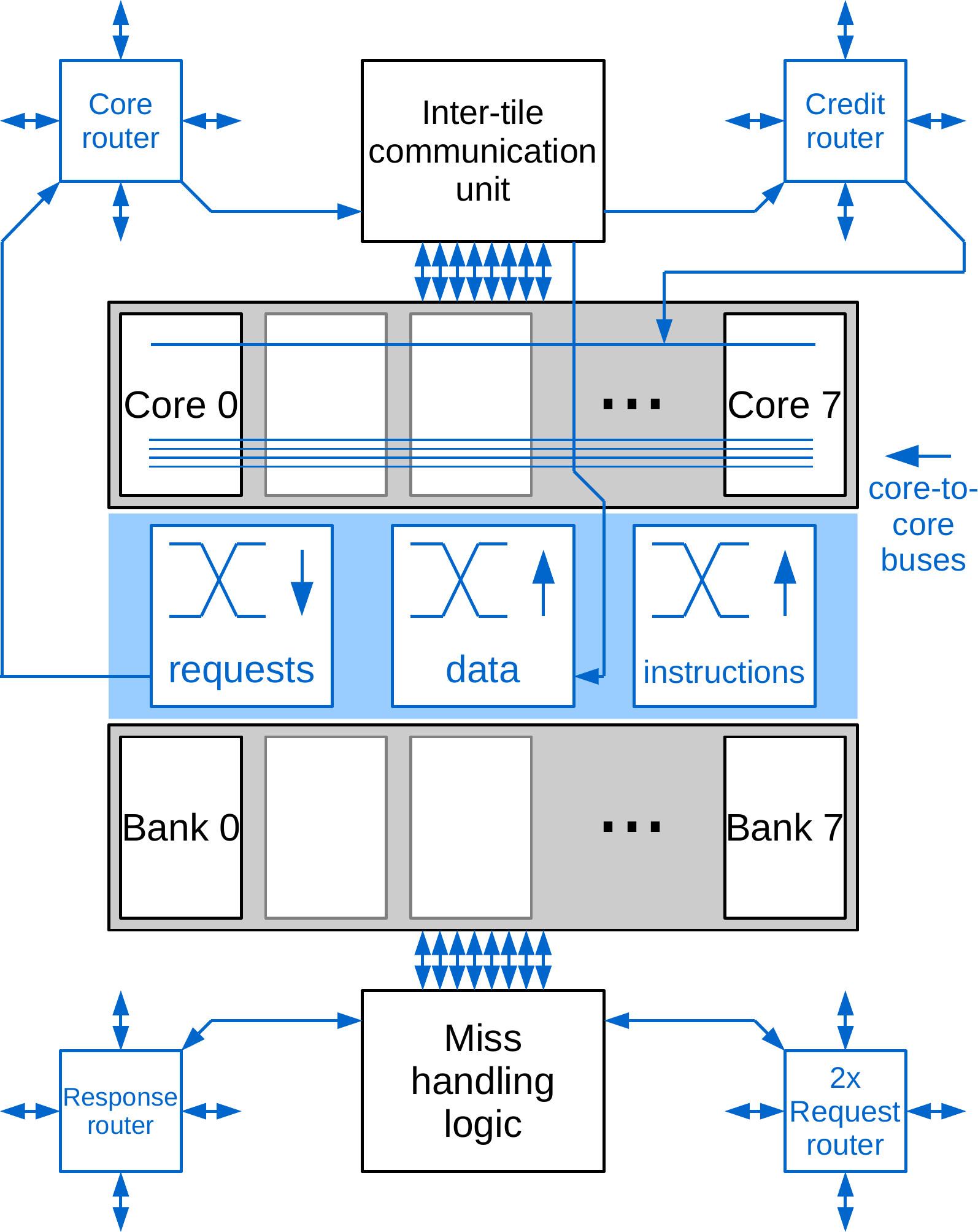}
  \caption{One tile of the Loki architecture.}
  \label{fig:tile}
\end{figure}

\begin{figure}
  \centering
  \includegraphics[width=\columnwidth]{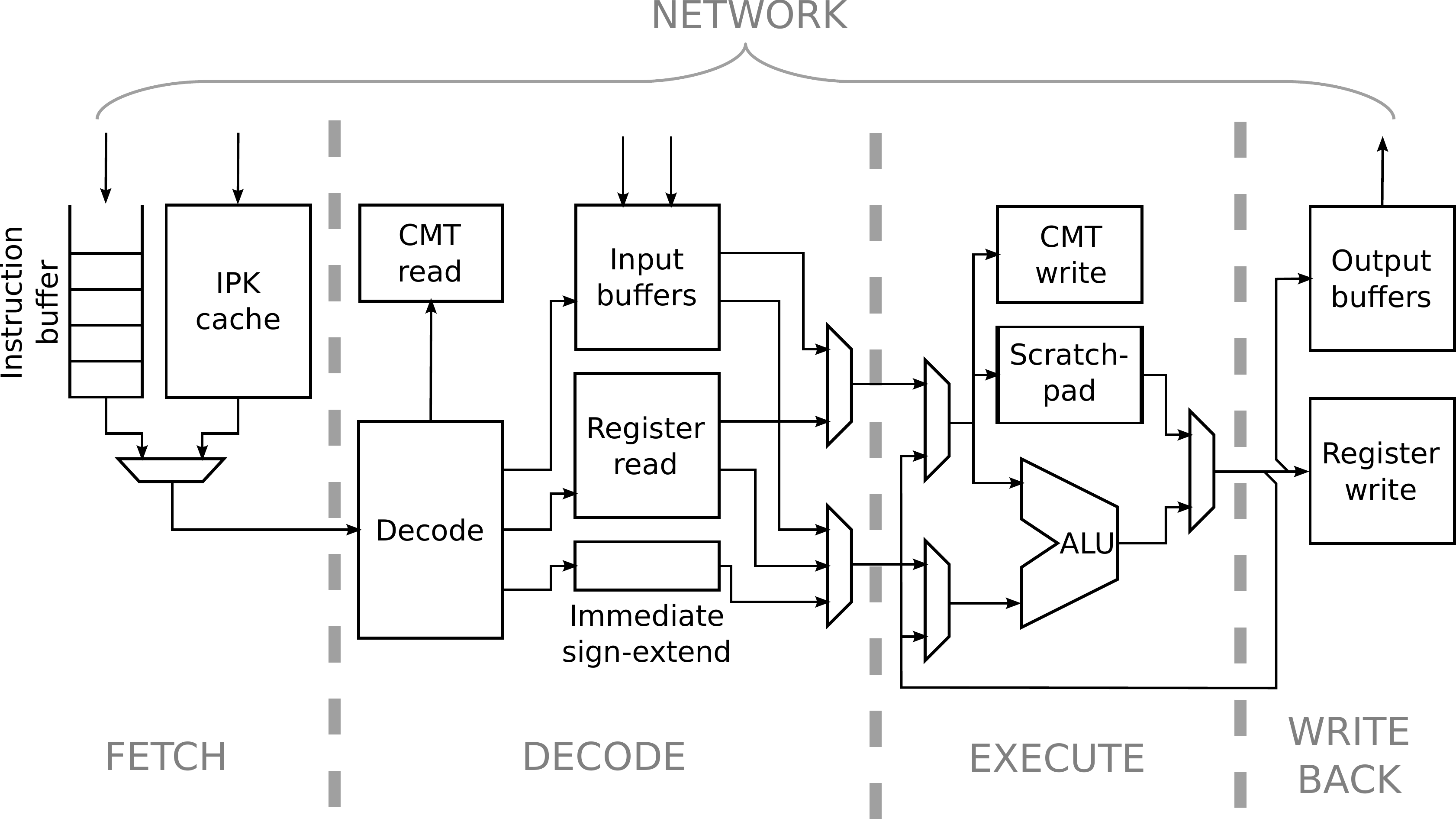}
  \caption{Loki core microarchitecture. IPK: instruction packet (Section~\ref{section:insts}), CMT: channel map table (Section~\ref{section:cmt}).}
  \label{fig:core}
\end{figure}

We use the Loki architecture as a baseline for our discussion \cite{lokijsps14}. Loki is a tiled architecture where each tile contains eight simple cores and eight memory banks (see Figures~\ref{fig:tile} and~\ref{fig:core}). A chip is made up of a large number of identical tiles, with each tile occupying approximately 1mm\textsuperscript{2} in a 40nm process. We plan to produce a modestly-sized test chip in 2016 containing 16 tiles (128 cores), so we restrict ourselves to this size for the rest of the paper.

Loki's design can be described as \emph{network-centric} -- communication between components is fast and efficient, allowing the boundaries between cores to be blurred. Communication within a tile takes a single clock cycle, and communication between routers on adjacent tiles also takes a single cycle. Network buffers are mapped to registers, allowing any instruction to read network data as though it was stored locally. Almost all instructions also allow their results to be sent over the network, in addition to being stored locally. Network communication is blocking: a component which attempts to read from an empty buffer will stall until data arrives, and a component which attempts to write to a full buffer will stall until space becomes available. All aspects of the network were designed with deadlock avoidance in mind: deadlock detection and recovery was deemed too complicated for such a simple architecture.

Both cores and memory banks can be accessed over the network, and it is possible for cores to send both instructions and data between each other. Sending of data allows the execution of multiple cores to be tightly-coupled. Additional cores can be used not just for parallel execution, but also to add resources (e.g. registers, memory bandwidth, network bandwidth) to another core. When used in this way, it may be possible to simplify cores to an extent where an individual core is rarely appropriate for executing a program, but resources can be added from other cores in a fine-grained way to suit the needs of each specific application. This would allow cores to be even smaller and more efficient. Sending of instructions between cores allows a range of interesting use-cases including inspection of remote state, rapid spawning of tasks and dynamic code rewriting.

Memory access also takes place over the network, with the memory request decoupled from the data access. A load instruction has no result, and simply sends a request to memory. Loki's ability to send the result of almost any operation over the network allows a rich variety of addressing modes. For example, the result of any bitwise operation can be treated as a memory address and sent to a memory bank to trigger the loading of data. This reduces the need for explicit load instructions, and goes some way towards reducing the impact of the relatively high memory latency (3 cycles: 1 cycle on the network to memory, 1 cycle memory access, 1 cycle on the network from memory). The memory bank returns data to one of the core's input buffers, to be accessed when the core is ready to do so. It is often possible to hide memory latency by scheduling independent instructions between the memory request and the data access.

In return for this relatively high L1 cache latency, Loki enjoys access to a larger cache than might otherwise be possible, since multiple memory banks are available to access. We find that different programs have different preferences: some are very sensitive to cache latency and do not suffer with a smaller capacity, while others are highly dependent on the amount of cache available and are able to tolerate higher latencies. Across a range of benchmarks, we found that there was not much difference between a smaller, faster cache and a larger, slower cache, so chose the latter as it allows for increased flexibility at runtime.

Loki contains a number of features which aid reconfiguration, and these are detailed in the following subsections. Our aim is to show that these features are justified and we hope to inspire the reader with possible ways in which a homogeneous architecture can be used to implement specialisation.

\subsection{Instruction transfer\label{section:insts}}
As mentioned previously, cores are able to send instructions to each other. Each core has two instruction channels to prevent these instructions from interfering with a program which is already executing. The primary channel connects to a small level-0 cache with a FIFO replacement policy. The secondary channel is typically used for these core-to-core communications and is uncached.

Instructions sent to a core could form a self-contained unit of work, such as inspecting the state of the target core and possibly returning a value, or the execution of the target core could be completely redirected with a request to begin executing from a given program counter. This feature allows additional state to be accessed quickly and efficiently, and also provides a mechanism for rapidly spawning new tasks or interrupting cores.

Loki's network-centric design lends itself to a packet-based instruction stream. Cores request instructions in atomic units which roughly correspond to basic blocks, and once a core begins execution of an instruction packet, it is guaranteed to continue to the end. Any scheduling decisions between the two instruction channels are made at the boundaries between instruction packets. If both channels hold instructions, the secondary channel has priority, otherwise instructions are executed from whichever channel has an instruction available.
    
\subsection{\texttt{\large sendconfig} instruction}
Loki supports an instruction called \texttt{sendconfig} which attaches arbitrary metadata to a payload. This is not expected to be used in regular execution, but proves itself useful in situations where low-level control over the hardware is desirable.

The metadata can describe information including the following:
\begin{itemize}
\item End-of-packet marker: wormhole routing is used throughout all networks, so once a packet starts being sent, network resources are reserved for it until the packet terminates.
\item Connection set-up/tear-down between remote cores.
\item Memory operation type.
\end{itemize}

The memory operation field, in particular, exposes operations which are usually only used within the memory system, such as invalidation, flushing and fetching of whole cache lines. Using these is much more efficient than requesting words individually, and can prevent unnecessary data movement. The network's packet system can be used when accessing memory to implement arbitrary atomic memory operations: a packet can be started with a request to load some data, and only terminated after some complex computation has been performed and a result written back. During this time, no other core will be able to access the same memory bank.
    
\subsection{Channel map table\label{section:cmt}}
\begin{figure}
  \centering
  \includegraphics[width=0.8\columnwidth]{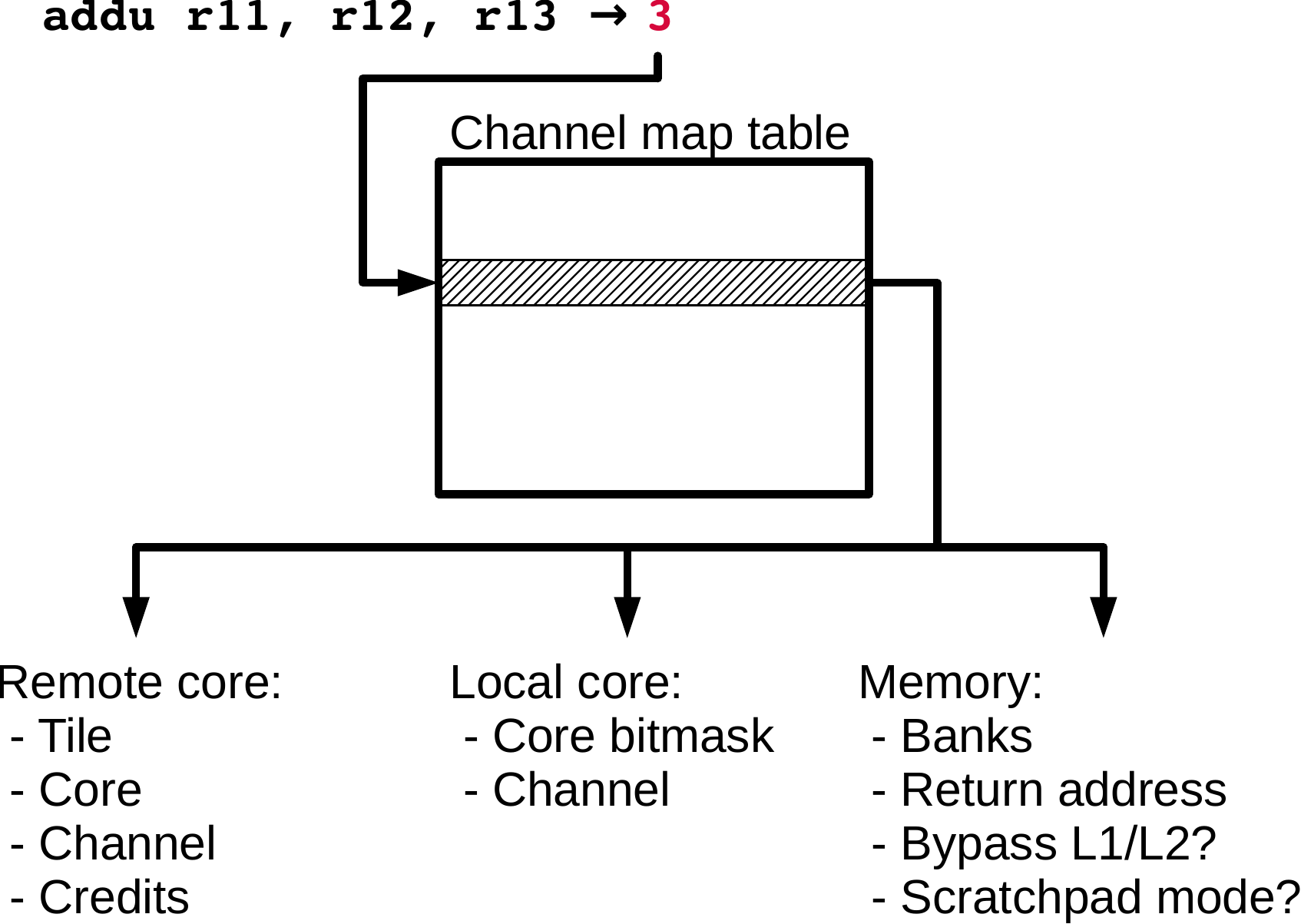}
  \caption{Channel map table access.}
  \label{fig:cmt}
\end{figure}

The channel map table maps logical network addresses to physical ones. Almost all instruction encodings include a field which allows the result of that instruction to be sent over the network in addition to being stored in a local register. Each core has its own channel map table which is consulted in the decode stage, at the same time as the register file is read. The channel map table contains rich information on which component should be contacted, and also the way in which it should be contacted (Figure~\ref{fig:cmt}).

A core on another tile is specified using a unique identifier, and the index of the input buffer to which data should be delivered. End-to-end flow control is used on these connections using a credit-based system, and the number of credits for each connection can be optionally configured in the channel map table. The default number of credits is the maximum number of spaces in the target buffer, as this ensures that any flit sent can be safely removed from the network at the other end. The credit count can be increased or decreased in software to control the rate of communication. Increasing the credit count is only safe if the target core is guaranteed to consume all data it receives, otherwise there is a risk of deadlock.

Cores on the local tile do not need to use credits since all networks within a tile are non-blocking and do not admit a new flit if it cannot be buffered at its destination. An arbitrary subset of cores on the local tile can be specified using a bitmask, and any data sent to them will be sent to all cores simultaneously.

Memory access provides the widest range of configuration options. Memory banks on the local tile can be combined into virtual groups, with cache lines distributed evenly across banks. Groups may overlap, and software is responsible for ensuring that all views of the same memory banks are consistent since no hardware coherence between banks is provided. Virtual memory groups can be used to provide dedicated storage space for different types of information: typically instructions and data, but a much finer granularity is possible, with separate space being provided for the stack and heap, or even individual data structures. Each group can be accessed in either cache or scratchpad mode, and each level of the memory hierarchy can be individually bypassed, allowing direct access to a shared L2 cache, or direct memory access, for example. It has previously been shown that scratchpad mode can be used frequently and can save around 10\% of total system energy by reducing the number of tag checks~\cite{andreasthesis}. Typically, memory banks will return information to the core that made the request, but it is possible to direct responses to another core if the software can guarantee that the target core will consume the data.

Updating the channel map table requires a single instruction and takes a single clock cycle. In many cases, the local core is the only component which needs to know about the update. The worst case is when reconfiguring the memory system in such a way that all dirty data must be written back to the next level of the memory hierarchy. This takes one clock cycle per word to be flushed back, plus one cycle per cache line.

\subsection{Level 2 cache directory}
In order to simplify Loki's implementation, L2 caches are built out of the same memory banks as L1 caches. Each tile is dynamically allocated as either a compute tile or an L2 tile. All 8 banks on an L2 tile are accessed in parallel to form an 8-way set-associative cache. Since the mapping of addresses to banks is not deterministic in a set-associative cache, the cores in an L2 tile cannot directly access the contents of their local memory banks. We are looking into lifting this restriction by dividing a tile's cache banks between L1 and L2 caches.

\begin{figure}
  \centering
  \includegraphics[width=0.7\columnwidth]{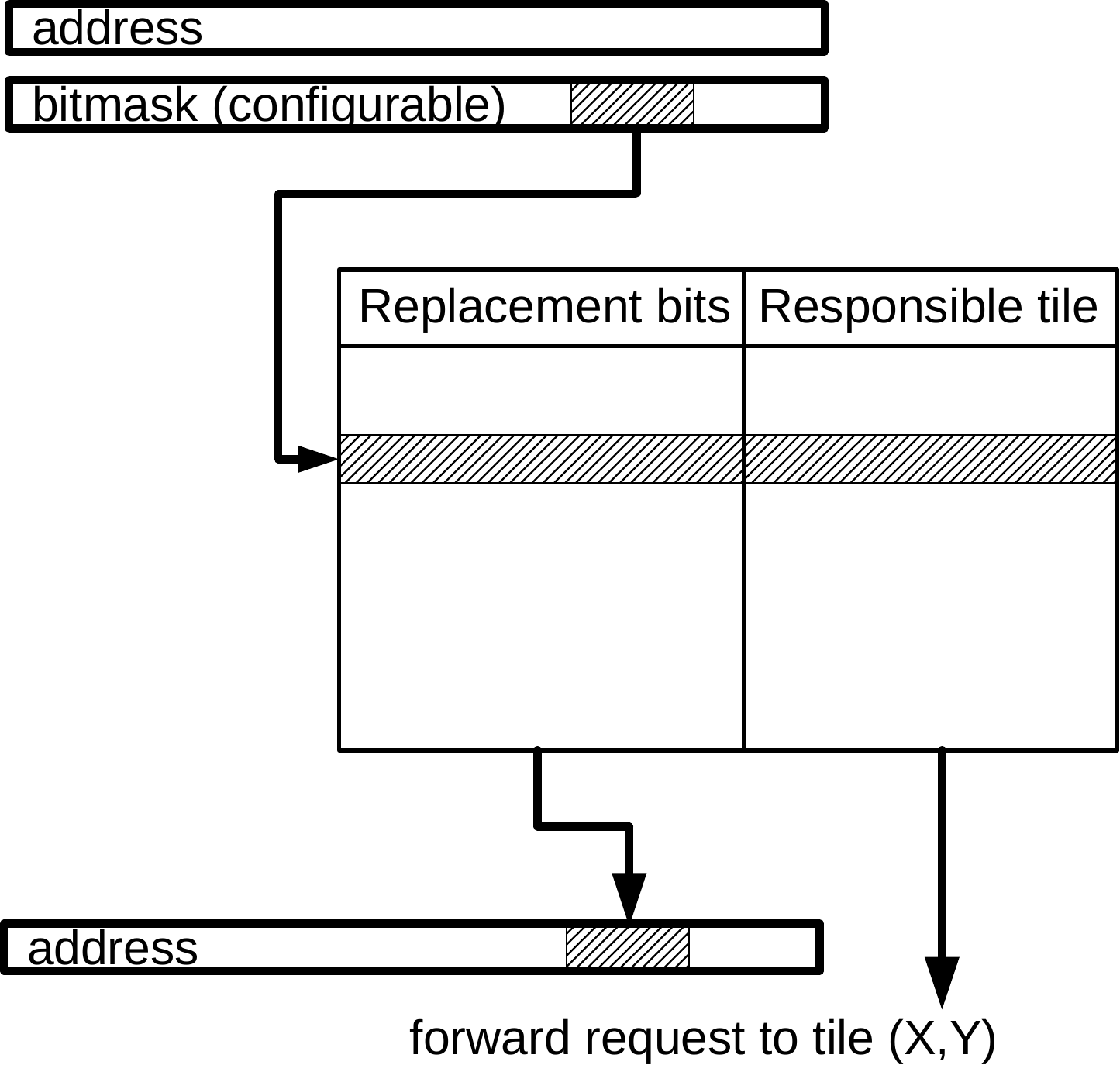}
  \caption{Memory directory access.}
  \label{fig:directory}
\end{figure}

Each tile contains a directory which is consulted after a cache miss (Figure~\ref{fig:directory}). The directory tells which tile is responsible for caching a given memory address, and also allows the substitution of a few bits of the address to allow a simple implementation of virtual memory. The target tile may be an L2 cache tile, or a memory controller which accesses off-chip memory. The directory is indexed using a few bits of the memory address which caused the cache miss. The position of these bits is reconfigurable: using less-significant bits tends to result in an even distribution of cache lines across L2 tiles, while using more-significant bits results in contiguous data being stored together with better locality. As with the L1 cache, it is possible to control which L2 tile(s) are responsible for storing particular information.

Having identical implementations for L1 and L2 memory banks means that both levels of the cache hierarchy are capable of high-bandwidth cache line operations and efficient sub-line accesses.

\section{Evaluation}
To evaluate the benefits of a configurable memory system, we use versions 1.1 and 2.0 of the EEMBC benchmark suite~\cite{eembc}. We use only those benchmarks which compile using the Loki toolchain~\cite{lokijsps14}, and only benchmarks which run to completion in a reasonable amount of time. This leaves benchmarks from all of the available sub-categories.

Benchmarks are executed on our SystemC simulator. The simulator has been validated against a SystemVerilog implementation, but runs much more quickly, allowing more interesting experiments to be performed. We instantiate a Loki architecture which is suited to these small benchmarks: memory banks are 2kB each, giving a total L1 cache capacity of 16kB per tile. Latency to main memory is set to 35 clock cycles -- a conservative figure which is likely to underestimate the penalty of a poorly-performing cache system, and therefore reduce any benefits observed by making better use of on-chip memory. Benchmarks are run once to warm the caches before any statistics are collected, and then execute repeatedly until all running benchmarks have completed at least one timed iteration.

Since the performance of the memory system correlates with the performance of running applications and with their energy efficiency (a higher hit rate means less data movement and fewer off-chip memory accesses), we only collect performance data in these experiments. We expect that any improvement in performance is very likely to be coupled with an improvement in energy consumption. We have previously performed extensive power modelling work for the Loki architecture~\cite{lokijsps14}.

\subsection{Single program}

\begin{figure*}
  \centering
  \includegraphics[width=\textwidth]{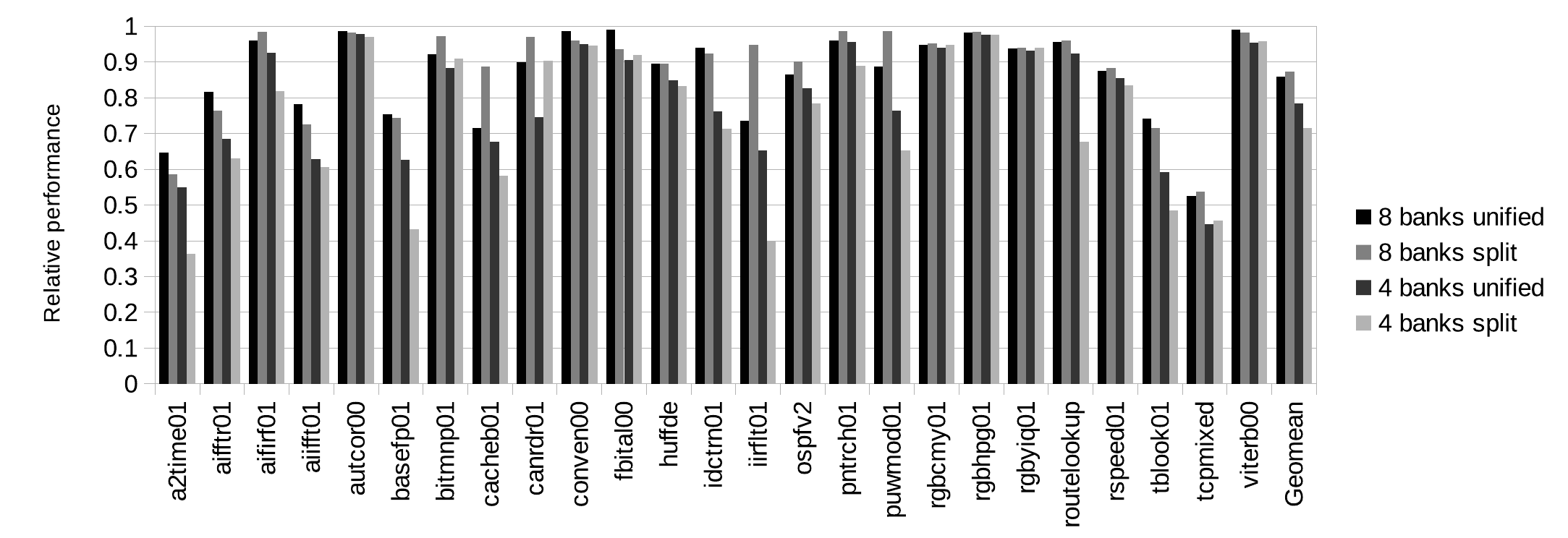}
  \caption{Performance of a selection of cache configurations, relative to an infinite-capacity, zero-latency L1 cache. Split caches are divided evenly between instructions and data.}
  \label{graph:l1best}
\end{figure*}

First, we run all benchmarks with all possible general L1 cache configurations and no L2 cache. We do not test any application-specific configurations, such as allocating memory banks to a particular data structure. We aim to find the best configuration to use for each benchmark when running in isolation, and to quantify the benefits of allowing configuration of the L1 cache.

Figure~\ref{graph:l1best} compares the performance of two 8-bank configurations and two 4-bank configurations. In each case, a unified cache and a cache split evenly between instructions and data is considered. With eight memory banks available, many benchmarks do not find the configuration to be very important -- the cache is used well with either configuration, or is inadequate with either configuration. For some, however, a large change is observed; \texttt{iirflt01} performs over 20\% worse with a unified cache than with a split cache. On average, since a configurable cache is able to select the best configuration for each application, a configurable L1 cache performs 3\% better than a fixed unified cache and 2\% better than a fixed split cache with eight memory banks.

This effect is magnified as cache contention increases. Extra contention could be due to larger applications having larger working sets, or from multiple applications needing to share the same memory banks. We can see the result of this by looking at the configurations which use fewer banks. With four banks, being able to select the best configuration gives an average 1\% improvement over a fixed unified cache and 10\% over a fixed split cache.

Again, \texttt{iirflt01} is the most variable benchmark, but interestingly, this time it prefers a unified cache, with performance almost 40\% worse when the cache is split. This seems to be because \texttt{iirflt01} is very sensitive to its instruction supply. When fewer memory banks are available, it prefers to have as many of them as possible capable of holding instructions. After about 4 banks, the instruction hit rate is close to 100\%, and it instead prefers to allocate additional banks to storing data, and splits the cache to avoid conflict misses.

Table~\ref{table:categories} shows how the different benchmarks behave with different amounts of cache. The `penalty' terms are somewhat subjective and take into account both how much and how quickly performance degrades when insufficient cache is available. Interestingly, there is little correlation between the number of banks required, and the penalty when fewer banks are provided. \texttt{fbital00}, for example, requires much more instruction cache than data cache. However, its performance is much more sensitive to the amount of data cache provided.

\begin{table}
\centering
\resizebox{\columnwidth}{!}{ 

\begin{tabular}{|l|c c|c c|} \hline
\multirow{2}{*}{Benchmark} & \multicolumn{2}{C{3cm}|}{Banks required for <10\% miss rate} & \multicolumn{2}{C{3.3cm}|}{Penalty when requirements not met}\\
 & Instructions & Data & Instructions & Data\\ \hline
\textbf{a2time01} & \textbf{>8} & \textbf{1} & \textbf{High} & \textbf{-}\\
\textbf{aifftr01} & \textbf{8} & \textbf{2} & \textbf{Moderate} & \textbf{Low}\\
\textbf{aifirf01} & \textbf{2} & \textbf{1} & \textbf{Extreme} & \textbf{-}\\
\textbf{aiifft01} & \textbf{8} & \textbf{2} & \textbf{Low} & \textbf{Low}\\
autcor00 & >8 & 1 & High & -\\
basefp01 & 4 & 1 & High & -\\
\textbf{bitmnp01} & \textbf{2} & \textbf{1} & \textbf{Low} & \textbf{-}\\
cacheb01 & 4 & >8 & High & Moderate\\
canrdr01 & 2 & 1 & Moderate & -\\
conven00 & 1 & 1 & - & -\\
fbital00 & 8 & 2 & Low & Extreme\\
huffde   & 1 & 2 & - & Moderate\\
idctrn01 & 4 & 1 & High & -\\
iirflt01 & 4 & 1 & Extreme & -\\
ospfv2   & 1 & 2 & - & Moderate\\
pntrch01 & 4 & 1 & High & -\\
puwmod01 & 4 & 4 & Low & Low\\
rgbcmy01 & 1 & 1 & - & -\\
rgbhpg01 & 1 & 1 & - & -\\
rgbyiq01 & 1 & 1 & - & -\\
routelookup & 4 & 4 & Low & Low\\
rspeed01 & 2 & 1 & Low & -\\
tblook01 & >8 & 1 & High & -\\
tcpmixed & >8 & 1 & Extreme & -\\
viterb00 & 1 & 1 & - & -\\ \hline
\end{tabular}

}
\caption{Benchmark sensitivities to different cache configurations. Bolded entries were selected for further examination in the following section.}
\label{table:categories}
\end{table}

\subsection{Multiprogramming}

\begin{figure*}
  \centering
  \includegraphics[width=\textwidth]{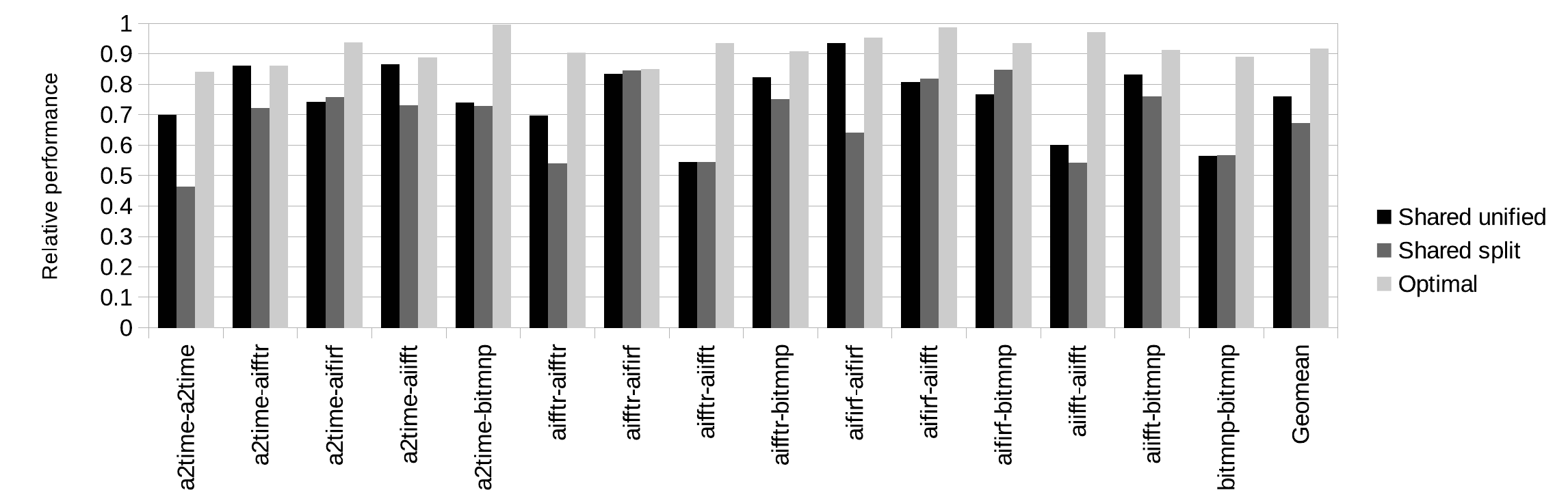}
  \caption{Performance of the best two-program L1 cache configuration and two fixed configurations, relative to the programs each running in isolation with access to all cache banks.}
  \label{graph:2programperformance}
\end{figure*}

We now explore how programs interact when competing for cache space. We select five benchmarks with various combinations of cache requirements and sensitivities, and then run each possible pair of benchmarks with each possible L1 cache configuration. There are a total of 64 configurations, with each cache also having the option of being shared between the two benchmarks or private. We scale each result relative to the best possible configuration when the application is running in isolation.

\begin{figure}
  \centering
  \includegraphics[width=\columnwidth]{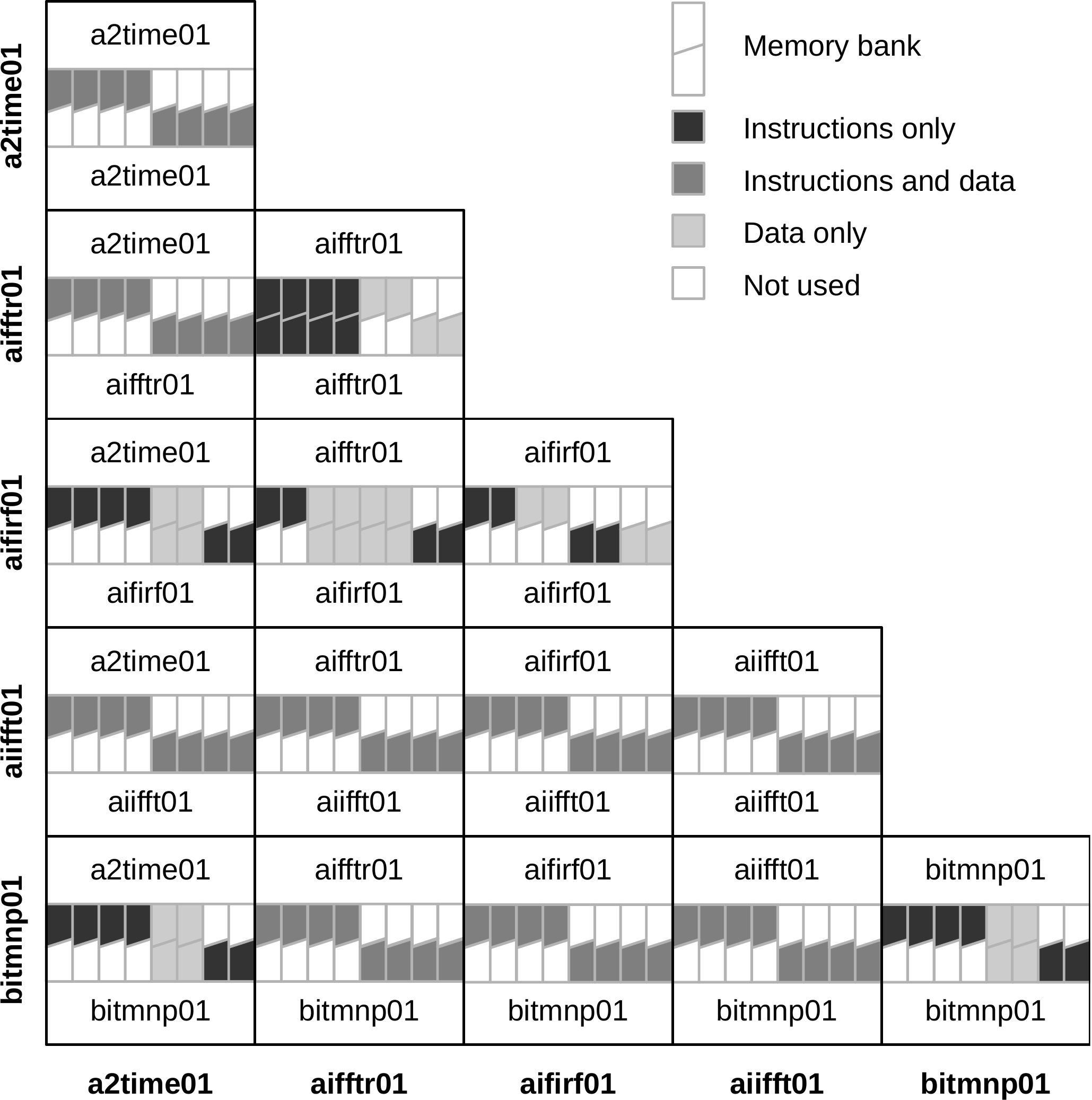}
  \caption{Optimal cache bank allocations for pairs of programs with equal priority.}
  \label{graph:2programallocation}
\end{figure}

Figure~\ref{graph:2programallocation} shows the best configuration for each pair of benchmarks when the performance of each benchmark is given an equal priority. (In practice, there is a range of optimal solutions in each case, depending on the relative priorities of the two applications.)

Ten of the fifteen combinations work best when the two programs are completely isolated from each other, with the remaining five performing best when some cache banks are shared. We also see a range of different decisions being made: some combinations prefer unified caches, some prefer split caches, some share the cache fairly between benchmarks, and some prefer to allocate more of the cache banks to one of the programs. Applications which are less sensitive to cache size are more likely to prefer a shared cache, but this is not always the case. \texttt{aiifft01}, for example, is not sensitive to either instruction or data cache capacity, yet always works best with private caches. This suggests that it is instead sensitive to cache latency, and prefers to have exclusive access to its memory banks to avoid contention.

Figure~\ref{graph:2programperformance} compares these configurations with the possible fixed configurations: unified and split shared caches. When sharing a tile's memory banks between two programs, each takes an average of 9\% longer compared to having all memory banks to itself. This compares very well with the 34\% (unified cache) and 49\% (split cache) slowdowns seen when using fixed configurations across all tests.

These performance improvements are due to better use of the available resources. The average L1 miss rate for the optimal cache configuration is reduced by a factor of 3 over the best fixed configuration. This results in considerably less off-chip communication, and less data movement in the on-chip memory, which leads to lower energy consumption.

The wide variety of optimal configurations for different situations shows that many degrees of freedom are necessary when configuring the memory system if maximum performance is to be achieved. The large gains over fixed configurations shows how much performance may be lost by more traditional architectures which are not able to implement application-specific configurations.

\subsection{Level 2 cache configuration}
\begin{figure}
  \centering
  \includegraphics[width=\columnwidth]{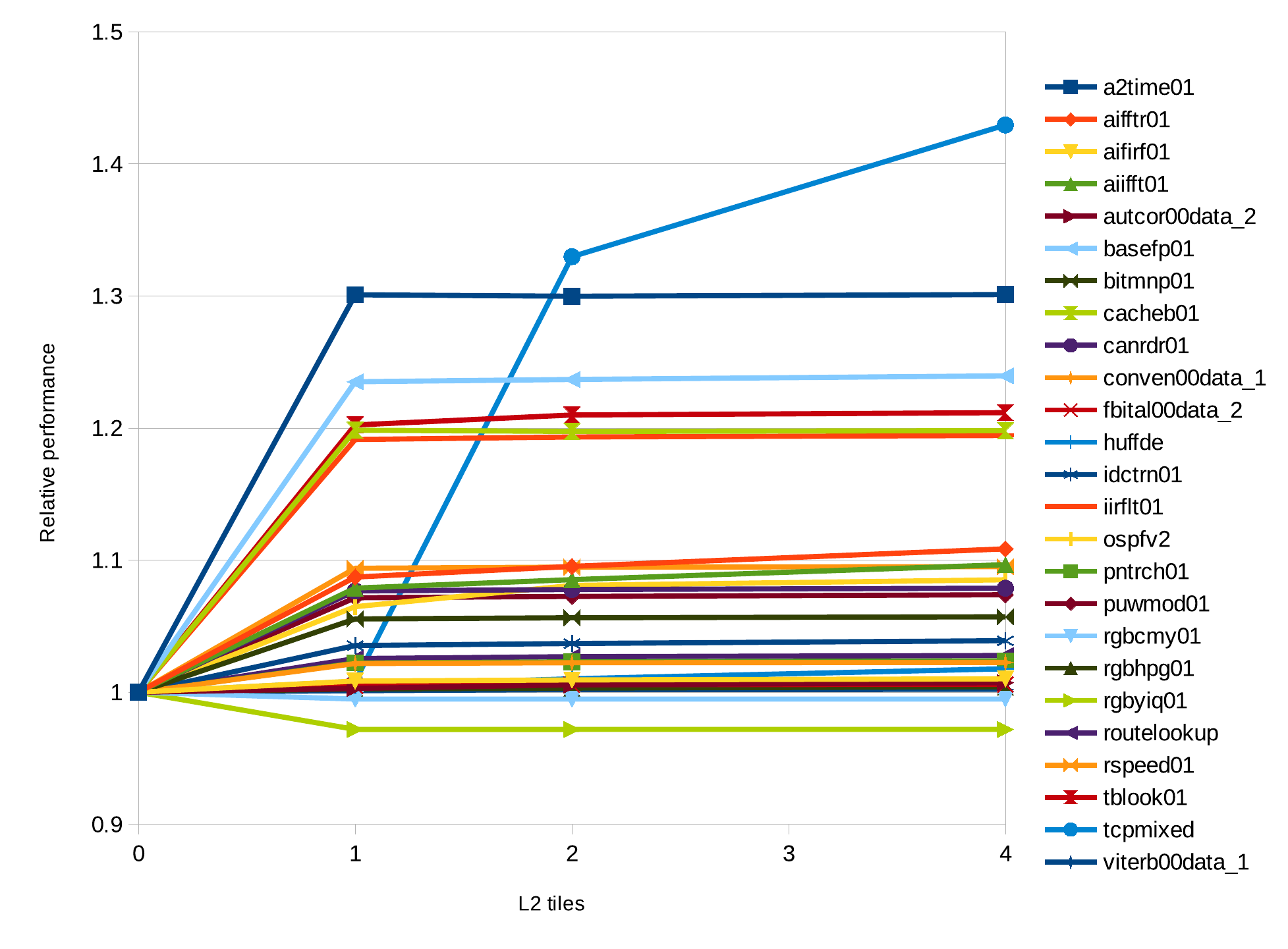}
  \caption{Performance with varying L2 cache size, relative to no L2 cache.}
  \label{graph:l2}
\end{figure}

We next look into changing the L2 cache. We give all benchmarks a unified L1 cache consisting of all 8 banks in the local tile, and vary the number of L2 tiles from 0 to 4. We ensure that all L2 tiles are adjacent to the compute tile. Other configurations would be possible and would simply introduce an extra cycle of latency in each direction for each additional tile traversed. Results are presented in Figure~\ref{graph:l2}.

As might be expected, different applications have different sensitivities to the varying L2 cache size. Most benchmarks exhibit a modest speedup; it would probably not be worth repurposing a computation tile as an L2 cache tile unless trying to accelerate a high-priority task, but if the tile was otherwise unused, it could be a simple and effective way to increase performance.

Some benchmarks perform worse when given an L2 cache. This is because they do not use any amount of cache effectively, and adding an extra level to the cache hierarchy simply adds to the latency when data is required from main memory. Conversely, Loki's flexibility allows these intermediate cache layers to be removed when they do not help, improving performance and allowing the resources to be allocated elsewhere.

The most interesting benchmark shown is \texttt{tcpmixed}. With a small L2 cache, performance is no better than with no L2 cache at all, indicating poor cache use. However, when the L2 cache increases in size enough to accommodate the working set of the benchmark, performance improves greatly.

Most of the benchmarks tested are not affected by L2 cache capacity beyond one tile. A flexible memory system allows the minimum necessary cache to be provided to each individual application, with the remaining resources being used elsewhere or power-gated to save energy.

\subsection{Case study}

\begin{table*}[t]
\centering
\begin{tabular}{ccccc} \hline
Architecture & Clock Frequency & Power & Area & Speed\\ \hline
One Loki core & $\sim$450MHz & $\sim$0.01W & 0.5mm$^2$ & 64.3 cycles/byte = 7MB/s\\
One Loki core + specialised L1 cache & $\sim$450MHz & $\sim$0.01W & 0.5mm$^2$ & 34.8 cycles/byte = 13MB/s\\
Eight Loki cores & $\sim$450MHz & $\sim$0.06W & 1mm$^2$ & 9.8 cycles/byte = 46MB/s\\
Eight Loki cores + specialised L1 cache & $\sim$450MHz & $\sim$0.06W & 1mm$^2$ & 6.5 cycles/byte = 69MB/s\\
One specialised Loki tile (as described) & $\sim$450MHz & $\sim$0.06W & 1mm$^2$ & 5.1 cycles/byte = 88MB/s\\
Sixteen specialised Loki tiles & $\sim$450MHz & $\sim$1W & 16mm$^2$ & 0.32 cycles/byte = 1411MB/s\\ \hline
One core of Intel\textregistered{} Core\texttrademark{} i7-980X \cite{aesinteli7} & 3.33 GHz & 33.5W & 60mm$^2$ & 1.3 cycles/byte = 2500MB/s\\
ARM9TDMI \cite{arm9tdmi,aesarm} & 150 MHz & 0.12W & 6.55mm$^2$ & 45 cycles/byte = 3.3MB/s\\
ATI Mobility Radeon\texttrademark{} HD 5650 \cite{aesradeon5650} & 650 MHz & 19W & 104mm$^2$ & 1.9 cycles/byte = 340MB/s\\
GeForce 8800 GTS \cite{aesgeforce8800} & 1.625 GHz & 135W & 484mm$^2$ & 17.1 cycles/byte = 95MB/s\\ \hline
\end{tabular}
\caption{AES implementations for a range of architectures.}
\label{table:aes}
\end{table*}

As a case study of the Loki architecture, AES-128-CTR mode encryption/decryption was implemented. This library was designed to use one whole tile, and have a network-centric API to allow easy embedding in a larger application. This would allow data to be streamed in or out of the AES library using message passing, or alternatively read and written directly from shared memory as a conventional AES hardware DMA block might. The application was designed, implemented, debugged and optimised in approximately two person-days by a programmer familiar with Loki's capabilities. A mixture of assembly language and C code was used -- it would be possible to have a pure C implementation, but our compiler is not yet able to optimise for many Loki-specific features.

The basic behaviour of AES-128 is to apply a round function 10 times on a block of data to generate the encrypted block. Each round, a fixed round key is used to modify the behaviour of the function. Mathematically, this could be represented as $f(k_9, f(k_8, ... f(k_1, f(k_0, data)) ... ))$. The last round should be a slightly different function to the other round functions according to the specification, but this can be re-expressed as a final correction function being applied after the final round, allowing all rounds to share a codebase. The size of the data block is always 128 bits, or 16 bytes. In CTR mode encryption and decryption, bulk data is encrypted by splitting the data into 16 byte blocks, and processing each separately.

\begin{figure}
  \centering
  \includegraphics[width=\columnwidth]{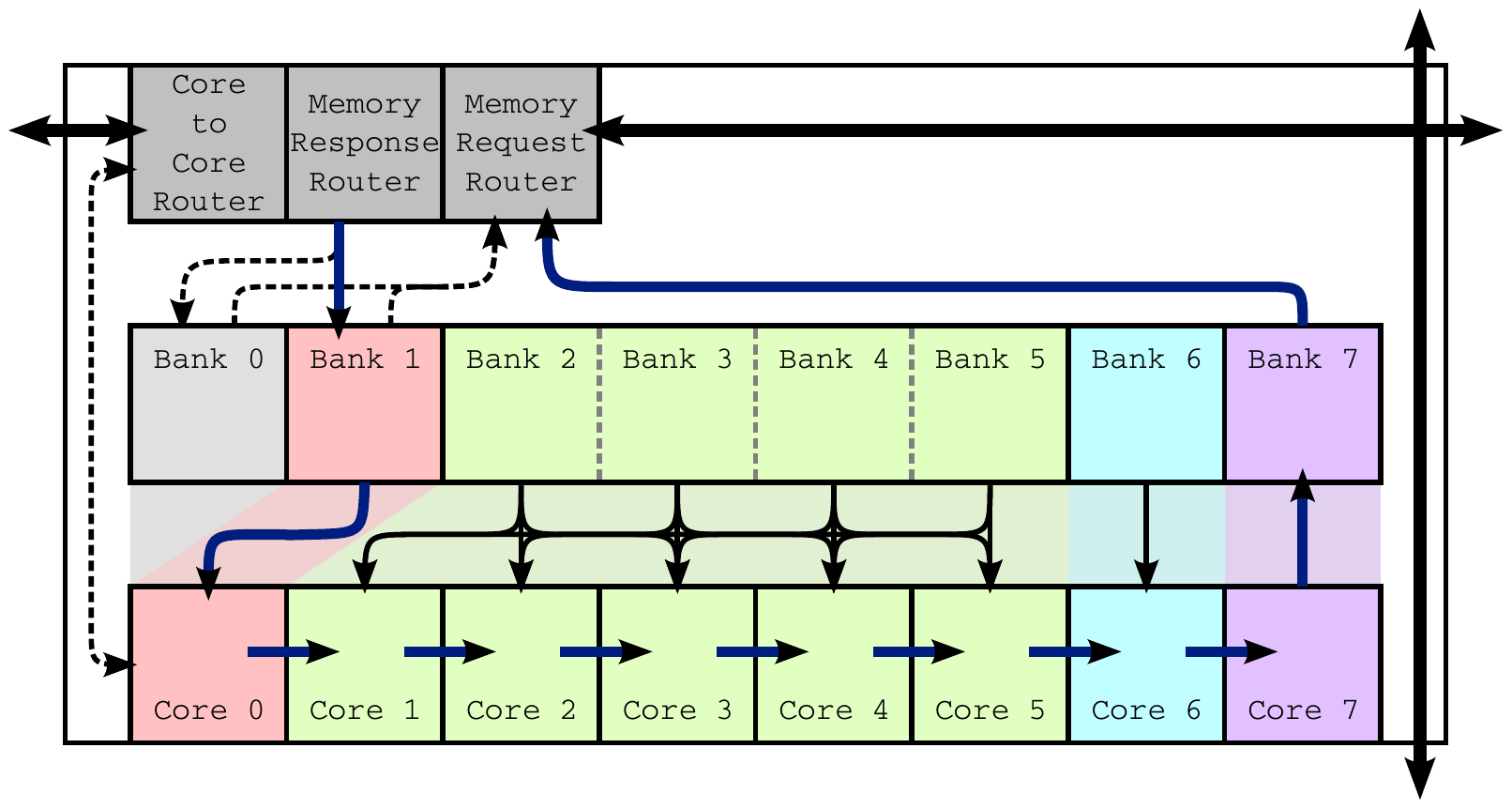}
  \caption{Resource allocation and communication patterns within the AES tile.}
  \label{fig:aes}
\end{figure}

The workload lends itself to several possible mappings to multiple Loki cores, for example processing the many independent blocks in parallel, exploiting ILP within each round function or forming a `software pipeline' in which each core applies two rounds and passes the data on to the next. A quick analysis suggested that this last mapping with the software pipeline would achieve the highest performance, so this was chosen and implemented (Figure~\ref{fig:aes}). Core 0 is given the responsibility of communicating with other tiles to provide the API for this library. Once a task is specified, this core starts the pipeline by preprocessing the data as required by CTR mode and sending it via message passing to core 1. Cores 1 to 5 all contain identical code to perform two rounds of AES on their input and pass it on. This takes 61 instructions, so fits in the 64-instruction L0 instruction cache, eliminating instruction fetch access to the L1 cache. The round keys can also be kept in registers, as each core must only remember two round keys, a total of 8 registers' worth of data. The final round correction function is then applied by core 6, which passes the data on to core 7. Core 7 is responsible for final output, either over the network or to memory.

The tile's memory banks are also specialised for various purposes. Bank 0 is left as a general purpose L1 instruction and data cache, mostly for convenience. It is not used at all during any of the tight loops. Bank 1 is dedicated to input memory operations as these will almost always result in cache misses, as input data is not frequently reused. By dedicating a bank to this, these misses never interfere with other memory requests. A \texttt{sendconfig} instruction is used to load a full cache line rather than issuing 8 individual loads. Another \texttt{sendconfig} instruction is used to prefetch the next cache line and avoid stalling the software pipeline on memory. Banks 2 to 5 are used as four individual lookup tables. These are used by the round functions on cores 1 to 5. These memory banks operate in scratchpad mode: their contents are pre-loaded, and it is known in advance that all required data fits in the banks. While the amount of data stored would fit entirely into one bank, using separate banks for each table simplifies the addressing and provides greater bandwidth. On average, each of these banks receives a load request every 2 clock cycles. Analysis of the interleaving of these loads allows us to compute the worst case response time of the banks, and thus schedule the reads to avoid stalls in the steady state interleaving. Bank 6 has a similar role, storing the tables needed by core 6's correction function. Finally, bank 7 has a similar role to bank 1, but for output rather than input. It is used exclusively for writeback of the results. This again makes use of \texttt{sendconfig} to store a whole cache line, thus avoiding have to first load the cache line as would be needed for 8 independent store instructions.

The ability for cores to send instructions to each other is used by core 0 to `rekey' the library. Whenever the encryption key changes (infrequently), the round keys are recomputed and then instructions are sent to each of the round function cores which change their round key registers appropriately. This allows the round function cores' code to be an infinite loop, saving a control flow instruction from this tight loop kernel.

The implementation is able to achieve a sustained performance of 5.1 cycles/byte on bulk encryption and decryption. This is quite an impressive figure, given that there is no architectural specialisation for AES. The implementation is also scalable to any number of tiles, by splitting the workload evenly between them all, which the library supports. Thus, with 6 or more tiles, we reach the point of multiple bytes per cycle.

Some figures published for other architectures are given in Table~\ref{table:aes}. The Intel processor achieves its impressive performance using the specialised AES-NI instruction set extension. When using a whole Loki chip, performance is comparable, while energy consumption and area are much lower.

Among the Loki implementations, we see that a highly-specialised memory configuration almost doubles performance in the single core case. This is partly due to improved cache behaviour, and partly due to fewer instructions being required to compute addresses when accessing scratchpads. Specialising the tile as a whole gives better results than simply repeating the single-core program on each core for two reasons:

\begin{enumerate}
\item There is less contention between cores trying to access the same memory bank at the same time.
\item Each core's task is much simpler, fitting in its private L0 cache, and reducing the strain on lower levels of the memory hierarchy. This is another example of specialisation allowing better utilisation of available resources.
\end{enumerate}

\subsection{Summary}
It has been shown that a configurable memory system can give significant gains in performance over a fixed memory system. Since performance gains in the memory system tend to correlate with a higher hit rate and therefore less data movement, we believe that large energy reductions are also possible. A side effect of improved cache performance is that execution time becomes more predictable -- this can be particularly useful in embedded systems which often must handle complex input and output data streams in real time.

A large number of different configurations perform best in different situations, many of which are impossible for an inflexible architecture to emulate. This shows the value in having a wide range of configuration options available. Adding configurability to a system does add an overhead, and will not be beneficial in every case, but we have found that the benefits outweigh the costs in the majority of cases.

\section{Discussion}  
We have seen in this paper how careful choice of memory configuration can have a large impact on performance. An architecture capable of dynamic reconfiguration is able to go further than this, however. Some applications may benefit from changing the memory configuration mid-execution to adapt to different phases of computation. For example, when entering its main loop, a program may benefit from reallocating some of its instruction banks to provide extra data capacity.

Finding which memory configuration is best can be a time-consuming task, however. It would be useful to extract information from a program, either statically or dynamically, and use it to predict which configuration would be best. Alternatively, a configuration could be selected, and then adjusted during execution based on performance metrics. We leave this automatic selection of configurations for future work.

While Loki is much more configurable than conventional architectures, some restrictions still remain. Lifting these remaining limitations could yield interesting results:

\begin{itemize}
\item Loki does not support a set-associative L1 cache, but some applications would certainly benefit from one. The L2 cache is capable of associativity by broadcasting requests to all banks on a tile; this is not possible in other situations as it adds too much complexity.
\item The size of virtual memory groups is restricted to powers-of-two to simplify the channel map table. In some cases, more control over bank allocation would be beneficial.
\item Loki guarantees deadlock freedom in its multi-level memory hierarchy by dedicating physically separate networks to different types of traffic. L1 $\rightarrow$ L2, L2 $\rightarrow$ main memory, and their responses all have their own networks. Deeper hierarchies on-chip are supported, but must be arranged in software to ensure that no messages of one type block the progress of messages of another type. In practice, this is straightforward: the lower levels of hierarchy should be placed physically closer to the off-chip memory controller.
\end{itemize}

\section{Related work}
Reconfiguration of a memory system is not a new idea: the Smart Memories architecture demonstrated that this could be a useful feature in the early 2000s~\cite{smart00}. Each cache line has four bits of metadata which can be manipulated by a configurable logic block. This allows more reconfiguration options including cache line sizes and replacement policies, but comes with a higher overhead. More recently, NVIDIA's Echelon project explored configurable memory hierarchies~\cite{echelon}. Each memory bank is given one or more `parent' banks, across which cache lines are evenly distributed. On a cache miss, data is requested from one of the parent banks. Neither of these works have examined their configuration spaces and potential benefits as we have here.

AsAP~\cite{asap09}, ACRES~\cite{acres04} and Raw~\cite{raw02} are all tiled architectures which exploit software specialisation to improve performance and energy efficiency, but none explore the memory system to the extent that we have here.

Various techniques have been proposed to reduce the number of cache conflicts and make cache access more predictable. These can involve pinning data in the cache to indicate that it should not be replaced~\cite{cachesetpinning} and partitioning the cache between different threads or cores~\cite{cachepartitioning,cacheqos} or different types of data~\cite{multilateralcache}, so they do not interfere with each other.

\section{Conclusions}
A large amount of the energy consumed by a modern embedded processor is in the memory system. By adding reconfigurability, it is possible for software to tailor the memory system to each application. This allows for less data movement and fewer accesses to large hardware structures, and improves both performance and energy consumption.

In this paper, we have explored the reconfiguration options of the Loki architecture's memory system. We found that despite its overheads, a reconfigurable system was worthwhile, and could significantly outperform any fixed implementation. Reconfigurability becomes more useful as resources become more constrained -- this is particularly relevant in embedded systems, where power constraints may limit the size of caches, while applications are getting increasingly complex and numerous. By adding reconfiguration options to a cache architecture, it is possible to make better use of the available cache, or to achieve the same performance with a smaller, more efficient cache.

\section{Acknowledgments}
We would like to thank the anonymous reviewers for their valuable feedback. This work was funded by the European Research Council grant number 306386.

%
\bibliographystyle{abbrv}
\bibliography{/homes/db434/Documents/Papers/loki}

\begin{thebibliography}{10}

\bibitem{aesinteli7}
K.~Akdemir, M.~Dixon, W.~Feghali, P.~Fay, V.~Gopal, J.~Guilford, E.~Ozturk,
  G.~Wolrich, and R.~Zohar.
\newblock Breakthrough {AES} performance with {Intel\textregistered{} AES New
  Instructions}.
\newblock White paper, 2010.

\bibitem{acres04}
B.~S. Ang and M.~Schlansker.
\newblock {ACRES} architecture and compilation.
\newblock Technical report, Hewlett-Packard, April 2004.

\bibitem{arm9tdmi}
{ARM Limited}.
\newblock {ARM922T} with {AHB} system-on-chip platform {OS} processor, 2007.

\bibitem{lokijsps14}
D.~Bates, A.~Bradbury, A.~Koltes, and R.~Mullins.
\newblock Exploiting tightly-coupled cores.
\newblock {\em Journal of Signal Processing Systems}, 80(1):103--120, 2015.

\bibitem{cachepartitioning}
J.~Chang and G.~S. Sohi.
\newblock Cooperative cache partitioning for chip multiprocessors.
\newblock In {\em Proceedings of the 21st Annual International Conference on
  Supercomputing}, ICS '07, pages 242--252, New York, NY, USA, 2007. ACM.

\bibitem{eembc}
{Embedded Microprocessor Benchmark Consortium and others}.
\newblock {EEMBC} benchmark suite.
\newblock http://www.eembc.org/, 2009.

\bibitem{multilateralcache}
A.~Gonz\'{a}lez, C.~Aliagas, and M.~Valero.
\newblock A data cache with multiple caching strategies tuned to different
  types of locality.
\newblock In {\em Proceedings of the 9th International Conference on
  Supercomputing}, ICS '95, pages 338--347, New York, NY, USA, 1995. ACM.

\bibitem{aesradeon5650}
S.~U. Haq, J.~Masood, A.~Majeed, and U.~Aziz.
\newblock Bulk encryption on {GPUs}.
\newblock White paper, October 2011.

\bibitem{cacheqos}
R.~Iyer, L.~Zhao, F.~Guo, R.~Illikkal, S.~Makineni, D.~Newell, Y.~Solihin,
  L.~Hsu, and S.~Reinhardt.
\newblock Qos policies and architecture for cache/memory in cmp platforms.
\newblock {\em SIGMETRICS Perform. Eval. Rev.}, 35(1):25--36, June 2007.

\bibitem{echelon}
S.~Keckler, W.~Dally, B.~Khailany, M.~Garland, and D.~Glasco.
\newblock {GPUs} and the future of parallel computing.
\newblock {\em Micro, IEEE}, 31(5):7--17, 2011.

\bibitem{andreasthesis}
A.~Koltes.
\newblock {\em Reconfigurable memory systems for embedded microprocessors}.
\newblock PhD thesis, University of Cambridge, 2014.

\bibitem{smart00}
K.~Mai, T.~Paaske, N.~Jayasena, R.~Ho, W.~J. Dally, and M.~Horowitz.
\newblock {Smart Memories}: a modular reconfigurable architecture.
\newblock In {\em Proceedings of the 27th annual International Aymposium on
  Computer Architecture}, ISCA '00, pages 161--171, New York, NY, USA, 2000.
  ACM.

\bibitem{aesgeforce8800}
H.~Nguyen.
\newblock {\em {GPU} Gems 3}.
\newblock Addison-Wesley, 2008.
\newblock Chapter 36. AES Encryption and Decryption on the GPU.

\bibitem{aesarm}
R.~Shamsuddin.
\newblock A comparative study of {AES} implementations on {ARM} processors.
\newblock Master's thesis, Oregon State University, 2005.

\bibitem{cachesetpinning}
S.~Srikantaiah, M.~Kandemir, and M.~J. Irwin.
\newblock Adaptive set pinning: Managing shared caches in chip multiprocessors.
\newblock {\em SIGARCH Comput. Archit. News}, 36(1):135--144, Mar. 2008.

\bibitem{raw02}
M.~B. Taylor, J.~Kim, J.~Miller, D.~Wentzlaff, F.~Ghodrat, B.~Greenwald,
  H.~Hoffman, P.~Johnson, J.-W. Lee, W.~Lee, A.~Ma, A.~Saraf, M.~Seneski,
  N.~Shnidman, V.~Strumpen, M.~Frank, S.~Amarasinghe, and A.~Agarwal.
\newblock The {Raw} microprocessor: A computational fabric for software
  circuits and general-purpose programs.
\newblock {\em IEEE Micro}, 22:25--35, March 2002.

\bibitem{asap09}
D.~Truong, W.~Cheng, T.~Mohsenin, Z.~Yu, A.~Jacobson, G.~Landge, M.~Meeuwsen,
  C.~Watnik, A.~Tran, Z.~Xiao, E.~Work, J.~Webb, P.~Mejia, and B.~Baas.
\newblock A 167-processor computational platform in 65 nm {CMOS}.
\newblock {\em Solid-State Circuits, IEEE Journal of}, 44(4):1130 --1144, April
  2009.

\end{thebibliography}
%

\end{document}